\documentclass[english,conference,final,a4paper]{IEEEtran}
 
\makeatletter

\usepackage[T1]{fontenc}
\usepackage[latin9]{inputenc}
\usepackage{amsthm}
\usepackage{amsmath}
\usepackage{amssymb}
\usepackage{graphicx}
\usepackage{dsfont}
\usepackage{cite}
\usepackage{amsmath}
\usepackage{array}
\usepackage{multirow}
\usepackage{caption}
\usepackage{color}

\theoremstyle{plain}

\theoremstyle{plain}
\newtheorem{lem}{\protect\lemmaname}
\theoremstyle{plain}
\newtheorem{thm}{\protect\theoremname}
\theoremstyle{plain}
  
\theoremstyle{definition}

\theoremstyle{definition}

\theoremstyle{definition}

\makeatother
  
\usepackage{babel} 
\providecommand{\lemmaname}{Lemma}
\providecommand{\propositionname}{Proposition}
\providecommand{\theoremname}{Theorem}
\providecommand{\corollaryname}{Corollary} 
\providecommand{\definitionname}{Definition}
\providecommand{\assumptionname}{Assumption}
\providecommand{\remarkname}{Remark}

\newcommand{\openone}{\mathds{1}}

\newcommand{\Hc}{\mathcal{H}}
\newcommand{\Ev}{\mathbf{E}}

\newcommand{\sign}{\mathrm{sign}}

\newcommand{\Binomial}{\mathrm{Binomial}}

\newcommand{\EE}{\mathbb{E}}
\newcommand{\PP}{\mathbb{P}}

\newcommand{\bsigma}{\boldsymbol{\sigma}}
\newcommand{\Poisson}{\mathrm{Poisson}}
\newcommand{\ktilde}{\tilde{k}}
 
\usepackage{color}

\begin{document}

\title{Partial Recovery Bounds for the \\ Sparse Stochastic Block Model}

\author{
 \IEEEauthorblockN{Jonathan Scarlett and Volkan Cevher}
  \IEEEauthorblockA{Laboratory for Information and Inference Systems (LIONS) \\
    \'Ecole Polytechnique F\'ed\'erale de Lausanne (EPFL) \\
    Email: \{jonathan.scarlett,volkan.cevher\}@epfl.ch}
} 

\maketitle

\begin{abstract}
    In this paper, we study the information-theoretic limits of community detection in the symmetric two-community stochastic block model, with intra-community and inter-community edge probabilities $\frac{a}{n}$ and $\frac{b}{n}$ respectively.  We consider the sparse setting, in which $a$ and $b$ do not scale with $n$, and provide upper and lower bounds on the proportion of community labels recovered on average.  We provide a numerical example for which the bounds are near-matching for moderate values of $a - b$, and matching in the limit as $a-b$ grows large. 
\end{abstract}

\long\def\symbolfootnote[#1]#2{\begingroup\def\thefootnote{\fnsymbol{footnote}}\footnote[#1]{#2}\endgroup}
\symbolfootnote[0]{ This work was supported in part by the European Commission under Grant ERC Future Proof, SNF 200021-146750 and SNF CRSII2-147633, and `EPFL Fellows' Horizon2020 grant 665667.}

\section{Introduction}

The problem of identifying community structures in undirected graphs is a fundamental problem in network analysis, machine learning, and computer science \cite{For10}, and is relevant to numerous practical applications such as social networks, recommendation systems, image processing, and biology.

The stochastic block model (SBM) is a widely-used statistical model for studying this problem.  Despite its simplicity, this model has helped to provide significant insight into the problem, has led to the development of several powerful community detection algorithms, and still comes with a variety of interesting open problems.

One such open problem, and the focus of the present paper, is to characterize the necessary and sufficient conditions for \emph{partial recovery}, in which one seeks to correctly recover a fixed proportion of the community assignments.  This is arguably of more practical interest compared to exact recovery, which is usually too stringent to be expected in practice, and compared to correlated recovery, which only seeks to marginally beat a random guess.

\subsection{The Symmetric Two-Community SBM}

We focus on the simplest SBM, in which there are only two communities and the edge probabilities are symmetric.  Specifically, the $n$ nodes, labeled $\{1,\dotsc,n\}$, are randomly assigned community labels $\bsigma = \{\sigma_1,\dotsc,\sigma_n\}$, where each $\sigma_i$ equals $1$ or $2$ with probability $\frac{1}{2}$ each.  Given the community labels, a set of ${n \choose 2}$ unordered edges $\Ev = \{E_{ij} \,:\, i \ne j\}$ is generated according to 
\begin{equation}
    \PP[E_{ij} = 1 \,|\, \bsigma] = 
    \begin{cases} 
        \frac{a}{n} & \sigma_i = \sigma_j \\  
        \frac{b}{n} & \sigma_i \ne \sigma_j,
    \end{cases}
\end{equation}
for some constants $a, b > 0$, with independence between different $(i,j)$ pairs.  We assume throughout the paper that $a$ and $b$ are fixed (i.e., not scaling with $n$), and hence the graph is sparse.  We also assume that $a > b$ (i.e., on average there are more intra-community edges than inter-community edges).  

Given the edge set $\Ev$, a decoder forms an estimate $\hat{\bsigma} := \{\hat{\sigma}_1,\dotsc,\hat{\sigma}_n\}$ of the communities.  Note that in this paper, we assume that $a$ and $b$ are known; this assumption is common in the literature, though sometimes avoided \cite{Abb15a,Gao15}.

\subsection{Previous Work and Contributions}

Studies of the SBM can roughly be categorized according to the recovery criteria of correlated recovery, exact recovery, and partial recovery.  A comprehensive review is not possible here, so we mention only some key relevant works.

The \emph{correlated recovery} problem only seeks to determine whether \emph{any} community structure is present or absent, thus insisting on classifying only a proportion $\frac{1}{2}(1+\epsilon)$ correctly for some arbitrarily small $\epsilon > 0$.  An exact phase transition between success and failure is known to occur according to whether $(a-b)^2 > 2(a+b)$ \cite{Mas14,Mos12}, as was conjectured in an earlier work based on tools from statistical physics \cite{Dec11}.

In the \emph{exact recovery} problem, one seeks to perfectly recover the two communities.  This is impossible with the above-mentioned scaling laws; instead, the main scaling regime of interest is $a,b = \Theta(\log n)$, in which a phase transition occurs according to whether $\frac{1}{\log n}\big(\frac{a+b}{2} - \sqrt{ab}\big) > 1$ \cite{Abb16}. Furthermore, this is achievable via practical methods \cite{Haj14,Abb16}, and extensions to the case of multiple communities and non-symmetric settings have been given \cite{Abb15}.

Several works have provided partial recovery bounds for the case that $a$ and $b$ exhibit certain scaling laws, or are finite but \emph{sufficiently large}.  In \cite{Mos13},  it is shown that a practical algorithm based on belief propagation achieves the optimal recovery proportion when $(a-b)^2 > C(a+b)$ for sufficiently large $C$. Bounds for several \emph{asymptotic} scalings of $a$ and $b$ are given in \cite{Gao15,And15,Mos15,Des15}, with \cite{And15,Gao15} considering a regime where the recovery proportion tends to zero, and \cite{Mos15,Des15} considering cases where the proportion tends to a constant.   A non-asymptotic bound is given in \cite{Gue14}, but the conditions on $a$ and $b$ are written in terms of a loose constant whose optimization is not attempted.  We are not aware of any previous works seeking tight performance bounds at finite values of $a$ and $b$.

In this paper, our goal is to partially close this gap by providing partial recovery bounds specifically targeted at the case that $a$ and $b$ are fixed and not necessarily large. We consider the partial recovery criterion
\begin{equation}
    r(\bsigma,\hat{\bsigma}) := \min_{\pi \in \Pi} \frac{1}{n}\sum_{i=1}^n \openone\big\{ \pi(\sigma_i) \ne \pi(\hat{\sigma}_i) \big\}, \label{eq:def_r}
\end{equation}
where $\Pi$ contains the two permutations of $\{1,2\}$; this is included since one can only hope to recover the communities up to relabeling.  Note that $r(\bsigma,\hat{\bsigma})$ is a random variable; we will primarily be interested in characterizing its expectation, but we will also present a high-probability bound.  

\subsection{Notation}

All logarithms have base $e$, and we define the binary entropy function in nats as $H_2(\alpha) := -\alpha\log\alpha - (1-\alpha)\log(1-\alpha)$.  The indicator function is denoted by $\openone\{\cdot\}$, and we use the standard asymptotic notations $O(\cdot)$, $o(\cdot)$, and $\Theta(\cdot)$.

\section{Main Results}

Here we present our main results, namely, information-theoretic bounds characterizing how the proportion of errors $r(\bsigma,\hat{\bsigma})$ can behave.  The proofs are given in Section \ref{sec:PROOFS}.  

\subsection{Necessary Condition}

We begin with a necessary condition that must hold for any decoding procedure.

\begin{thm} \label{thm:necc}
    \emph{(Necessary Condition)}
    Under the symmetric SBM with fixed parameters $a > b > 0$, any decoder must yield
    \begin{equation}
        \liminf_{n\to\infty} \EE\big[r(\bsigma,\hat{\bsigma})\big] \ge \PP\big[ Z_1 < Z_2 \big] + \frac{1}{2}\PP\big[ Z_1 = Z_2 \big], \label{eq:necc}
    \end{equation}
    where $Z_1 \sim \Poisson\big( \frac{a}{2} \big)$, $Z_2 \sim \Poisson\big( \frac{b}{2} \big)$ are independent.
\end{thm}

The proof is based on a \emph{global to local} relation from \cite{And15}, roughly stating that the best average error rate is equal to the best average error rate in estimating a single assignment (node 1, say).  Assuming the best case scenario that all other nodes are estimated correctly, the estimation of the remaining node roughly amounts to performing a Poisson hypothesis test \cite{Abb15}, thus yielding the expression in \eqref{eq:necc} in terms of Poisson random variables.

\subsection{Sufficient Conditions}

Next, we provide our sufficient conditions.  Note that these are purely information-theoretic, as the decoders used in the proofs are not computationally feasible.  We first provide a \emph{high probability} bound based on a minimum-bisection decoder, which has also been considered in previous works such as \cite{Abb16}.  We will see that this bound is reasonable but sometimes loose; nevertheless, it will provide the starting point for an improved bound given in Theorem \ref{thm:suff} below.

\begin{thm} \label{thm:suff_hp}
    \emph{(High-Probability Sufficient Condition)}
    Under the symmetric SBM with fixed parameters $a > b > 0$, there exists a decoder such that, for any $\epsilon > 0$, there exists $\psi > 0$ such that
    \begin{equation}
        \PP[ r(\bsigma,\hat{\bsigma}) > \alpha+\epsilon ] \le e^{-\psi n} + \frac{1}{n^2}, \label{eq:high_prob}
    \end{equation}
    for sufficiently large $n$, where $\alpha\in\big(0,\frac{1}{2}\big]$ is defined to be the solution to 
        \begin{equation}
            \frac{a+b}{2} - \sqrt{ab} = \frac{ H_2(\alpha) }{ \alpha(1 - \alpha) } \label{eq:alpha_def}
        \end{equation}
        if such a solution exists, and $\alpha = 0.5$ otherwise.
\end{thm}

Our main sufficient condition is given as follows.

\begin{thm} \label{thm:suff}
    \emph{(Refined Sufficient Condition)}
    Under the symmetric SBM with fixed parameters $a > b > 0$, suppose that there exists a value $\alpha\in\big(0,\frac{1}{4}\big)$ satisfying \eqref{eq:alpha_def}.  Then there exists a decoding procedure such that
    \begin{equation}
        \limsup_{n\to\infty} \EE\big[r(\bsigma,\hat{\bsigma})\big] \le \PP\big[ Z_{1,\alpha} < Z_{2,\alpha} \big] + \frac{1}{2}\PP\big[ Z_{1,\alpha} = Z_{2,\alpha} \big], \label{eq:suff}
    \end{equation}
    where $Z_{1,\alpha} \sim \Poisson\big( \frac{a}{2}(1-\alpha) + \frac{b}{2}\alpha \big)$ and $Z_{2,\alpha} \sim \Poisson\big( \frac{b}{2}(1-\alpha) + \frac{a}{2} \alpha \big)$ are independent.
\end{thm}

The proof uses a two-step decoding procedure inspired by \cite{Gao15}, in which the first step uses the decoder from Theorem \ref{thm:suff_hp}, and the second step performs local refinements.  We again liken this to a Poisson-based testing procedure to obtain \eqref{eq:suff}.  Note that this condition takes a similar form to that in \eqref{eq:necc}; we will see numerically in Section \ref{sec:NUMERICAL} that the gap between the two is often small, particularly when $a-b$ is large.

\subsection{Discussion and a Conjectured Sufficient Condition} \label{sec:CONJECTURE}

The proof of our main achievability bound, Theorem \ref{thm:suff}, is based on using a high probability bound in the first step, and then obtaining an improved bound in the second step using local refinements.  If we could show that the average-distortion bound in Theorem \ref{thm:suff} also holds with high probability (e.g., $1-o\big(\frac{1}{n}\big)$), then we could use this overall procedure in the first step of a new two-step procedure, and then obtain a further improved bound of the form \eqref{eq:suff}, with our current achievability \eqref{eq:suff} bound playing the role of $\alpha$.  

One could then imagine repeating this argument several times, further improving the bound on each iteration.  See Section \ref{sec:NUMERICAL} for a numerical example.

Even if this argument can be formalized, there is still a major hurdle in handling small values of $a-b$: We require an initial high probability bound with a fraction of errors strictly smaller than $\frac{1}{4}$.  Theorem \ref{thm:suff_hp} does not suffice for this purpose in general, and refined methods for obtaining such bounds would be of significant interest.  Alternatively, one could seek to adjust the two-step procedure so that one may start with a high probability bound considering \emph{any} fraction of errors in $\big(0,\frac{1}{2}\big)$, rather than just $\big(0,\frac{1}{4}\big)$.

\subsection{Numerical Example} \label{sec:NUMERICAL}

\begin{figure} 
    \begin{centering}
        \includegraphics[width=1\columnwidth]{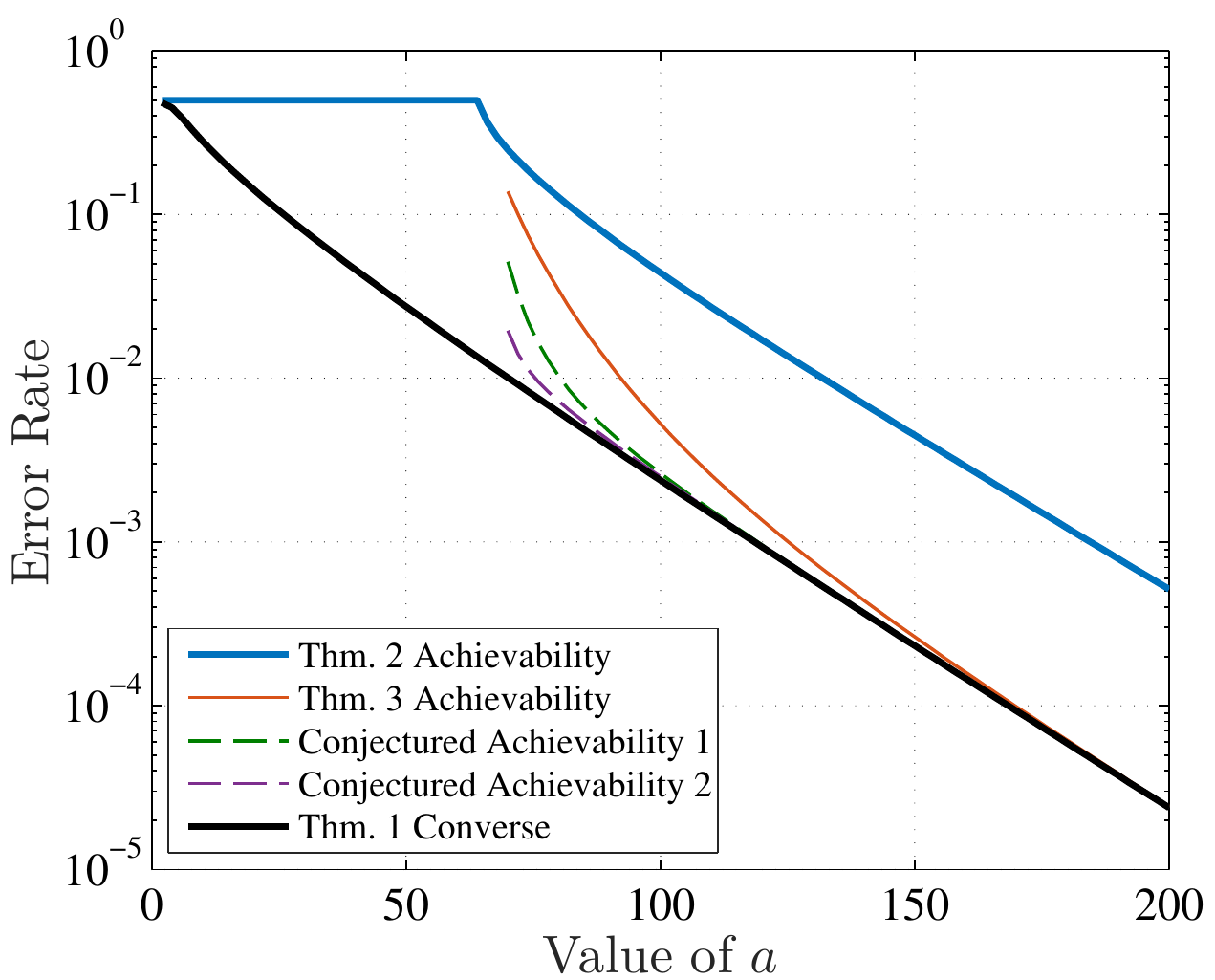}
        \par
    \end{centering}
    
    \caption{Asymptotic partial recovery bounds with $a = 2b$. The vertical axis gives the limit of $\EE[r(\bsigma,\hat{\bsigma})]$ as $n \to \infty$. \label{fig:example}}
    \vspace*{-4ex}
\end{figure}

In Figure \ref{fig:example}, we plot our asymptotic bounds for various values of $(a,b)$ such that $a = 2b$.  Thus, higher values of $a$ (or equivalently, $b$) correspond to a larger gap between $a$ and $b$, making the community detection problem easier.  Along with the main achievability and converse bounds, we plot the high probability achievability bound (i.e., the solution to \eqref{eq:alpha_def}).  Moreover, we plot the bounds that would arise from the first two iterations of the iterative procedure corresponding to the conjectured sufficient condition described in Section \ref{sec:CONJECTURE}.

While the high probability bound provides a similar rate of decay to the converse bound as $a$ increases, the gap between the two at finite values of $a$ remains significant.  In contrast, our main achievability bound from the two-step procedure approaches the converse bound as $a \to \infty$, which is to be expected since this procedure bears similarity to the asymptotically optimal two-step procedure proposed in \cite{Gao15}.

In contrast, our bounds have more room for improvement at low values of $a$.  In particular, results from the correlated recovery problem \cite{Mas14,Mos12} reveal that one can achieve an error rate better than $\frac{1}{2}$ if and only if $(a-b)^2 > 2(a+b)$, or equivalently $a > 12$ (since we are considering the case $a = 2b$).  Our converse bound is below $\frac{1}{2}$ for all $a>0$, our high-probability achievability bound is still equal to $\frac{1}{2}$ for $a=60$, and our refined achievability bound is only valid for $a \gtrsim 70$, since it relies on the high-probability bound being below $\frac{1}{4}$.  

Closing these gaps for small values of $a$ and $b$ is a challenging but interesting direction for future work.  While our conjectured sufficient condition appears that it could help significantly at moderate values of $a$ and $b$, it still has the same limitations when these values are small.  The techniques of \cite{Mos15} may also be useful, since the genie argument used in the converse part is more general than the one we use, and the belief propagation decoder used in the achievability part is potentially more powerful at small values of $a$ and $b$.

\section{Proofs} \label{sec:PROOFS}

Here we provide the proofs of Theorems \ref{thm:necc}--\ref{thm:suff}.  Due to space constraints, we omit some details that are in common with previous works such as \cite{Abb16} and \cite{And15}.

\subsection{Proof of Necessary Condition (Theorem \ref{thm:necc})} \label{sec:PF_NECC}

The proof is based on a \emph{global to local} lemma given in \cite{And15}.  Recall that $\Pi$ is the set of permutations of $\{1,2\}$ corresponding to reassignments (of which there are only two, since we consider the two-community case), and define $S(\bsigma,\hat{\bsigma}) = \{ \bsigma' \,:\, \bsigma' = \pi(\hat{\bsigma}),  r(\bsigma,\hat{\bsigma}) = \frac{1}{n}\sum_{i=1}^n \openone\{ \pi(\sigma_i) \ne \pi(\hat{\sigma}_i), \pi \in \Pi \}  \big\}$, containing the reassignments of $\hat{\bsigma}$ corresponding to the set of permutations achieving the minimum in \eqref{eq:def_r} (typically a singleton).  

\begin{lem} \emph{(Global to local \cite{And15})}
    The minimum value of $\EE[r(\bsigma,\hat{\bsigma})]$ over all decoders is equal to the minimum value of  $\EE\big[\frac{1}{|S(\bsigma,\hat{\bsigma})|} \sum_{\bsigma' \in S(\bsigma,\hat{\bsigma})} \openone\{\sigma_1 \ne \sigma'_1\}\big]$ over all decoders.
\end{lem}

This result essentially allows us to obtain a lower bound on the error rate $\EE[r(\bsigma,\hat{\bsigma})]$ via a lower bound on the error rate corresponding to the first node.  For the latter, we consider a genie-aided setting in which the true assignments of nodes $2,\dotsc,n$ are revealed to the decoder, which is left to estimate node $1$.  We can then assume without loss of optimality that $\hat{\sigma}_i = \sigma_i$ for $i=2,\dotsc,n$, and in this case we have $S(\bsigma,\hat{\bsigma}) = \{\hat{\bsigma}\}$.  Thus, we are left to bound $\EE\big[\frac{1}{|S(\bsigma,\hat{\bsigma})|} \sum_{\bsigma' \in S(\bsigma,\hat{\bsigma})} \openone\{\sigma_1 \ne \sigma'_1\}\big] = \PP[\sigma_1 \ne \hat{\sigma}_1]$.  Note that the information from the genie only makes the recovery of $\sigma_1$ easier, and hence any converse bound for this setting is also valid for the original setting.

Suppose that, among the revealed nodes $2,\dotsc,n$, there are $n_1 := \frac{n-1}{2}(1+\delta)$ nodes in community 1, and $n_2 := \frac{n-1}{2}(1-\delta)$ in community 2, for some $\delta \in [-1,1]$.  Since the community assignments are independent and equiprobable, Hoeffding's inequality \cite[Ch.~2]{Bou13} gives the following with probability at least $1-\frac{1}{n^2}$:
\begin{equation}
    |\delta| \le 2\sqrt{ \frac{\log n}{n-1} }. \label{eq:delta_bound}
\end{equation}
For fixed $\delta$, the study of the error event $\{\sigma_1 \ne \sigma'_1\}$ in the genie-aided setting comes down to a binary hypothesis testing problem, where hypothesis $\Hc_\nu$ ($\nu=1,2$) is that $\sigma_1 = \nu$.  Letting $\ell_\nu$ denote the number of edges from node 1 to nodes from $2,\dotsc,n$ that are in the $\nu$-th community, we have
\begin{align}
    \Hc_1 ~:~ &\ell_1 \sim \Binomial\Big( \frac{n-1}{2}(1+\delta), \frac{a}{n} \Big), \nonumber \\
        & \ell_2 \sim \Binomial\Big( \frac{n-1}{2}(1-\delta), \frac{b}{n} \Big) 
\end{align}
\vspace*{-3ex}
\begin{align}
    \Hc_2 ~:~ &\ell_1 \sim \Binomial\Big( \frac{n-1}{2}(1+\delta), \frac{b}{n} \Big), \nonumber \\
        & \ell_2 \sim \Binomial\Big( \frac{n-1}{2}(1-\delta), \frac{a}{n} \Big).
\end{align}
We now observe, as in \cite{Abb15}, that this problem can be approximated by a Poisson hypothesis testing problem of the form
\begin{align}
    &\Hc'_1 \,:\, \ell_1 \sim \Poisson\Big( \frac{a}{2}(1+\delta) \Big), \,\ell_2 \sim \Poisson\Big( \frac{b}{2}(1-\delta) \Big) \label{eq:H1'} \\
    &\Hc'_2 \,:\, \ell_1 \sim \Poisson\Big( \frac{b}{2}(1+\delta) \Big), \,\ell_2 \sim \Poisson\Big( \frac{a}{2}(1-\delta) \Big). \label{eq:H2'}
\end{align}
Specifically, we have from Le Cam's inequality \cite[Eq.~(32)]{Abb15} that each Binomial distribution above differs from the corresponding Poisson distribution by $O\big(\frac{1}{n}\big)$ in the total-variation norm, and hence the difference in the error rates resulting from the two hypothesis testing problems is also $O\big(\frac{1}{n}\big)$.

Recalling that our hypotheses are equiprobable, a substitution of the Poisson probability mass function (PMF) $p_k = \frac{\lambda^k}{k!} e^{-\lambda}$ into \eqref{eq:H1'}--\eqref{eq:H2'} reveals that the decision rule minimizing the error rate is to choose $\Hc'_1$ if and only if
\begin{equation}
    \ell_1 \ge \ell_2 + \frac{\delta(b-a)}{ \log\frac{a}{b} }.
\end{equation}
Using \eqref{eq:delta_bound} and the fact that $a$ and $b$ do not scale with $n$, we find that $\big|\frac{\delta(b-a)}{ \log\frac{a}{b} }\big| < 1$ for sufficiently large $n$, and hence the decision simply amounts to testing which of $\ell_1$ and $\ell_2$ is larger, with ties broken according to whether $\delta$ is positive (choose $\Hc'_1$), negative (choose $\Hc'_2$), or zero (choose randomly).  For example, under $\Hc'_1$ with $\delta = 0$, we find that the probability of incorrectly choosing $\Hc'_2$ is
\begin{equation}
    \PP[ Z'_1 < Z'_2 ] + \frac{1}{2}\PP[ Z'_1 = Z'_2 ], \label{eq:necc_final_delta}
\end{equation}
where $Z'_1 \sim \Poisson\big( \frac{a}{2}(1+\delta) \big)$ and $Z'_2 \sim \Poisson\big( \frac{b}{2}(1-\delta) \big)$.  Since $\delta \to 0$ by \eqref{eq:delta_bound}, the error rate in \eqref{eq:necc_final_delta} approaches that given in \eqref{eq:necc}.  By handling the other cases of $\Hc$ and $\sign(\delta)$ similarly, we find that the overall error rate also approaches the right-hand side of \eqref{eq:necc}, thus completing the proof.

\subsection{Proof of High-Probability Sufficient Condition (Theorem \ref{thm:suff_hp})} \label{sec:PF_SUFF_HP}

The theorem is trivial for $\alpha = \frac{1}{2}$, since even a random guess recovers half of the communities correctly on average; we thus focus on the case that $\alpha \in \big(0,\frac{1}{2}\big)$.  We also assume that $n$ is even; otherwise, the same result follows by simply ignoring an arbitrary node and assigning its community at random.

We consider a minimum-bisection decoder that splits the $n$ nodes into two communities of size $\frac{n}{2}$, such that the number of inter-community connections is minimized.  This decoder was studied in several previous works such as \cite{Abb16,And15}.

We begin by conditioning on the true community assignments having $n_1 = \frac{n}{2}(1+\delta)$ nodes in community 1, and $n_2 = \frac{n}{2}(1-\delta)$ nodes in community 2.  As we showed in the converse proof, we have with probability at least $1-\frac{1}{n^2}$ that $\delta$ satisfies \eqref{eq:delta_bound}; this is what leads to the second term in \eqref{eq:high_prob}.

\begin{figure}
    \begin{centering}
        \includegraphics[width=0.7\columnwidth]{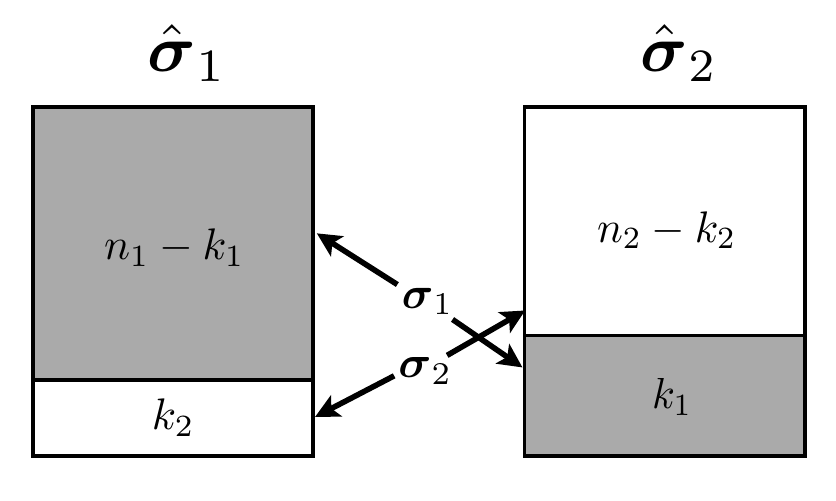}
        \par
    \end{centering}
    
    \caption{Sizes of true communities and their estimates in the case that $\delta > 0$ (i.e., $n_1 > n_2$). \label{fig:estimates}} \vspace*{-4ex}
\end{figure}

Consider a fixed estimate $\hat{\bsigma}$ of the communities from the above procedure, and suppose that there are $k_\nu$ indices such that $\sigma_i = \nu$ but $\hat{\sigma}_i \ne \nu$ ($\nu=1,2$).  See Figure \ref{fig:estimates} for an illustration.  Since the decoder always declares exactly $\frac{n}{2}$ nodes to be in each of the two communities, we must have $\frac{n}{2}(1+\delta) - k_1 + k_2 = \frac{n}{2}$ and $\frac{n}{2}(1-\delta) - k_2 + k_1 = \frac{n}{2}$, and hence $k_1 - k_2 = \frac{n}{2}\delta$ or equivalently $k_1 + k_2 = 2k_2 + \frac{n}{2}\delta$.  Since $k_1 + k_2$ corresponds to the total number of mis-labeled communities, and since $\delta$ satisfies \eqref{eq:delta_bound}, in order to have $r(\bsigma,\hat{\bsigma}) > \alpha(1+\eta)$, it is necessary that $k_2 > \frac{n}{2}\alpha$ and $k_2 < \frac{n}{2}(1-\alpha)$ for sufficiently large $n$ (recall from \eqref{eq:def_r} that the recovery is only defined up to relabeling).

We now consider the probability that a fixed estimate yielding some $(k_1,k_2)$ pair is chosen by the minimum-bisection decoder.  We focus on the case that $k_2 \in \big(\frac{n}{2}\alpha,\frac{n}{4}\big]$ and $k_1 \le k_2$ (i.e., $\delta > 0$), since the cases with $k_2 \in \big[\frac{n}{4}, \frac{n}{2}(1-\alpha)\big]$ or $k_2 > k_1$ are handled analogously.  In order for an error to occur, the true assignment must yield a lower number of inter-community connections than the assignment obtained by swapping $k_1$ incorrect nodes from community 1 with $k_1$ incorrect nodes from community 2.  Such a swap causes $k_1(n_1 - k_1) + k_1(n_2 - k_2) = k_1(n - k_1 - k_2)$ inter-community edges to have probability $\frac{b}{n}$ instead of $\frac{a}{n}$, as well as $k_1(k_1 - k_2) = k_1\frac{n}{2}\delta$ inter-community edges to have probability $\frac{a}{n}$ instead of $\frac{b}{n}$.  Thus, in order for an error occur, a random variable of the following form (corresponding to the inter-community edges differing in the two assignments) must be non-negative:
\begin{equation}
    \Psi_{k_1,k_2} := W_{1,b} - W_{1,a} + W_{2,a} - W_{2,b},
\end{equation}
where $W_{1,a} \sim \Binomial\big(k_1(n - k_1 - k_2), \frac{a}{n}\big)$ and $W_{2,a} \sim \Binomial\big(k_1\frac{n}{2}\delta,\frac{a}{n}\big)$, and analogously for $W_{1,b}$ and $W_{2,b}$ with $b$ in place of $a$.

Applying the union bound and a simple counting argument, we obtain
\begin{equation}
    \PP[\mathrm{error} \,|\, \bsigma] \le 2 \sum_{k_2 = \frac{n}{2}\alpha}^{\frac{n}{4}} { n_1 \choose k_1 }{ n_2 \choose k_2 } \PP[ \Psi_{k_1,k_2} > 0 ], \label{eq:union_bound}
\end{equation}
where $k_1 = k_2 + \frac{n}{2}\delta$, and $\bsigma$ is an arbitrary assignment with $n_\nu$ nodes in community $\nu$ ($\nu=1,2$).  The factor of $2$ here arises from a symmetry argument with respect to the estimates with $k_2 < \frac{n}{4}$ and $k_2 > \frac{n}{4}$.

Let $P_A$ and $P_B$ denote Bernoulli PMFs with parameters $\frac{a}{n}$ and $\frac{b}{n}$, respectively.  An application of the Chernoff bound yields for any $\lambda > 0 $ that
\begin{multline}
    \PP[\Psi_{k_1,k_2} > 0] \le \bigg( \sum_{z_a,z_b} P_A(z_a)P_B(z_b) e^{\lambda(z_b-z_a)} \bigg)^{m_1} \\ \times\bigg( \sum_{z_a,z_b} P_A(z_a)P_B(z_b) e^{\lambda(z_a-z_b)} \bigg)^{m_2}, \label{eq:chernoff}
\end{multline}
where $m_1 := k_1(n - k_1 - k_2)$, $m_2 := k_1\frac{n}{2}\delta$, and $z_a,z_b \in \{0,1\}$.  It is straightforward to show that the choice of $\lambda$ minimizing the first summation is $\lambda = \frac{1}{2}\log\frac{ \frac{a}{n}(1-\frac{b}{n}) }{ \frac{b}{n}(1-\frac{a}{n}) }$, and that the summation evaluates to $2\sqrt{ \frac{a}{n}(1-\frac{b}{n}) \frac{b}{n}(1-\frac{a}{n}) } + \frac{a}{n}\frac{b}{n} + \big(1-\frac{a}{n}\big)\big(1-\frac{b}{n}\big)$.  The second summation also behaves as $1+\Theta\big(\frac{1}{n}\big)$, and since $m_1 = \Theta(n^2)$ but $m_2 = o( n^2 )$, we obtain the following after applying some asymptotic expansions:
\begin{equation}
   \hspace*{-1.5ex}-\frac{1}{n} \log \PP[\Psi_{k_1,k_2} > 0] \ge \frac{m_1}{n^2} \bigg( 2\Big( \frac{a+b}{2} - \sqrt{ab} \Big) \bigg) + o(1). \label{eq:chernoff2}
\end{equation}
Supposing now that $k_2 = \frac{n}{2}\alpha_0$ for some $\alpha_0\in\big[\alpha,\frac{1}{2}\big]$ (see \eqref{eq:union_bound}), we readily obtain from \eqref{eq:delta_bound} that $k_1 = \frac{n}{2}\alpha_0 (1+o(1))$ and $m_1 = \frac{1}{2}n^2 \alpha_0(1-\alpha_0) (1+o(1))$, and we similarly have $n_1 = \frac{n}{2} (1+o(1))$ and $n_2 = \frac{n}{2}(1+o(1))$.  Substituting these estimates and \eqref{eq:chernoff2} into  \eqref{eq:union_bound} and using the identity $\frac{1}{N}\log{N \choose \theta N} = H_2(\theta) (1+o(1))$, we find that the right-hand side of \eqref{eq:union_bound} decays to zero exponentially fast provided that
\begin{equation}
    H_2(\alpha_0) - \alpha_0(1-\alpha_0)\Big( \frac{a+b}{2} - \sqrt{ab} \Big) < 0
\end{equation}
for all $\alpha_0 \in\big[\alpha,\frac{1}{2}\big]$.  Since $\frac{H_2(\alpha_0)}{\alpha_0(1-\alpha_0)}$ is monotonically decreasing in this range, this holds provided that $\alpha$ satisfies \eqref{eq:alpha_def}.

\subsection{Proof of Refined Sufficient Condition (Theorem \ref{thm:suff})} \label{sec:PF_SUFF}

We again assume that $n$ is even, and the case that $n$ is odd follows similarly by ignoring one node and assigning its community randomly.  Theorem \ref{thm:suff_hp} allows us to prove Theorem \ref{thm:suff} via the following two-step procedure  \cite{Gao15}:
\begin{enumerate}
    \item For each $j=1,\dotsc,n$, do the following:
     \begin{enumerate}
         \item Apply the decoder from Theorem \ref{thm:suff_hp} to the set of nodes $\{1,\dotsc,n\} \backslash \{j\}$ to obtain the estimates $\{\tilde{\sigma}_i^{(j)}\}_{i \ne j}$.  Choose the remaining estimate $\tilde{\sigma}^{(j)}_j$ in such a way that there are an equal number of nodes with $\tilde{\sigma}^{(j)}_j = 1$ and $\tilde{\sigma}^{(j)}_j = 2$.
         \item If there are more values of $i$ with $\tilde{\sigma}_i^{(j)} = \tilde{\sigma}_i^{(1)}$ than $\tilde{\sigma}_i^{(j)} \ne \tilde{\sigma}_i^{(1)}$, set each $\hat{\sigma}_i^{(j)} = \tilde{\sigma}_i^{(j)}$.  Otherwise, set each $\hat{\sigma}_i^{(j)}$ to be the value differing from $\tilde{\sigma}_i^{(j)}$.
     \end{enumerate}
    \item For each $j=1,\dotsc,n$, set the final estimate $\hat{\sigma}_j = 1$ if there are more edges from node $j$ to nodes with $\hat{\sigma}_i^{(j)} = 1$ than to nodes with $\hat{\sigma}_i^{(j)} = 2$, and set $\hat{\sigma}_j = 2$ otherwise.
\end{enumerate}

We again write $n_1 = \frac{n}{2}(1+\delta)$ and $n_2 = \frac{n}{2}(1-\delta)$, and note that $\delta$ satisfies \eqref{eq:delta_bound} with probability at least $1-\frac{1}{n^2}$. 

Let $\alpha'$ be an arbitrary value in the range $\big(\alpha,\frac{1}{4}\big)$.  For each $j=1,\dotsc,n$, let $\ktilde^{(j)}_{\nu}$ ($\nu = 1,2$) be the number of nodes from the $\nu$-th community such that the $j$-th decoder in Step 1 outputs $\tilde{\sigma}_i^{(j)} \ne \nu$, and let $k^{(j)}_{\nu}$ be defined similarly with  $\hat{\sigma}_i^{(j)}$ in place of $\tilde{\sigma}_i^{(j)}$.  By Theorem \ref{thm:suff_hp} and the union bound, with probability $1 - O\big(\frac{1}{n}\big)$, we have for all $j$ that either $\ktilde^{(j)}_1 + \ktilde^{(j)}_2 \le n\alpha'$ or $\ktilde^{(j)}_1 + \ktilde^{(j)}_2 \ge n(1-\alpha')$.

We consider the case that $\ktilde^{(1)}_1 + \ktilde^{(1)}_2 \le n\alpha'$; the other case  $\ktilde^{(1)}_1 + \ktilde^{(1)}_2 \ge n(1-\alpha')$ is handled analogously.  From the above definitions and Step 1b above, we trivially have $k^{(1)}_{\nu} = \ktilde^{(1)}_{\nu}$, and hence $k^{(1)}_1 + k^{(1)}_2 \le n\alpha'$.  We claim that it is also the case that $k^{(j)}_1 + k^{(j)}_2 \le n\alpha$ for $j=2,\dotsc,n$.  Indeed, since $\alpha' < \frac{1}{4}$, the contrary would imply that less than a quarter of the $\hat{\sigma}_i^{(1)}$ differ from the true assignments and more than three quarters of the $\hat{\sigma}_i^{(j)}$ differ from the true assignments, in turn implying that more than half of the $\hat{\sigma}_i^{(1)}$ differ from the $\hat{\sigma}_i^{(j)}$, in contradiction with Step 1b above.

By definition, among the $\hat{\sigma}_i^{(j)}$, there are $\frac{n}{2}(1+\delta) - k_1^{(j)} + k_2^{(j)}$ nodes estimated to be in community $1$, and $\frac{n}{2}(1-\delta) - k_2^{(j)} + k_1^{(j)}$ to be in community $2$.  Since the decoder from Step 1 outputs an estimate with an equal number $\frac{n}{2}$ of nodes in each community, this implies  that $k_1^{(j)} - k_2^{(j)} = \frac{n}{2}\delta$.  Summing this with $k_1^{(j)} + k_2^{(j)} \le n\alpha'$, we obtain $k_1^{(j)} \le \frac{n}{2}\big(\alpha' + \frac{\delta}{2}\big)$, and subtracting the two equations similarly gives $k_2^{(j)} \le \frac{n}{2}\big(\alpha' - \frac{\delta}{2}\big)$.

Finally, we consider the testing procedure given in Step 2 above.  We have the following when $\sigma_j = 1$: (i) To nodes with $\hat{\sigma}_i^{(j)} = 1$ there are $n_1 - k_1^{(j)}$ potential edges having probability $a$ and $k_2^{(j)}$ having probability $b$; (ii) To nodes with $\hat{\sigma}_i^{(j)} = 2$ there are $n_2 - k_2^{(j)}$ potential edges having probability $b$ and $k_1^{(j)}$ having probability $a$.  When $\sigma_j = 2$, the same is true with the roles of $a$ and $b$ reversed.

The proof is now completed in the same way as Section \ref{sec:PF_NECC} by approximating each of these numbers of edges by a Poisson distribution.  The above estimates, along with \eqref{eq:delta_bound}, reveal that $n_1$ and $n_2$ behave as $\frac{n}{2}  + o(n)$, and each $k_\nu^{(j)}$ is upper bounded by $\frac{n}{2}\alpha' + o(n)$.  In the worst case scenario that these upper bounds are met with equality, the parameters of the resulting Poisson distributions converge to $\frac{a}{2}(1-\alpha') + \frac{b}{2}\alpha'$ and $\frac{b}{2}(1-\alpha') + \frac{a}{2}\alpha'$.  Since $\alpha'$ can be chosen to be arbitrarily close to $\alpha$, this leads to the final bound given in \eqref{eq:suff}.

\vspace*{-0.5ex}
\bibliographystyle{IEEEtran}
\bibliography{JS_References}

\begin{thebibliography}{10}
\providecommand{\url}[1]{#1}
\csname url@samestyle\endcsname
\providecommand{\newblock}{\relax}
\providecommand{\bibinfo}[2]{#2}
\providecommand{\BIBentrySTDinterwordspacing}{\spaceskip=0pt\relax}
\providecommand{\BIBentryALTinterwordstretchfactor}{4}
\providecommand{\BIBentryALTinterwordspacing}{\spaceskip=\fontdimen2\font plus
\BIBentryALTinterwordstretchfactor\fontdimen3\font minus
  \fontdimen4\font\relax}
\providecommand{\BIBforeignlanguage}[2]{{%
\expandafter\ifx\csname l@#1\endcsname\relax
\typeout{** WARNING: IEEEtran.bst: No hyphenation pattern has been}%
\typeout{** loaded for the language `#1'. Using the pattern for}%
\typeout{** the default language instead.}%
\else
\language=\csname l@#1\endcsname
\fi
#2}}
\providecommand{\BIBdecl}{\relax}
\BIBdecl

\bibitem{For10}
S.~Fortunato, ``Community detection in graphs,'' \emph{Physics Reports}, vol.
  486, no.~3, pp. 75--174, 2010.

\bibitem{Abb15a}
E.~Abbe and C.~Sandon, ``Recovering communities in the general stochastic block
  model without knowing the parameters,'' 2015,
  http://arxiv.org/abs/1506.03729.

\bibitem{Gao15}
C.~Gao, Z.~Ma, A.~Y. Zhang, and H.~H. Zhou, ``Achieving optimal
  misclassification proportion in stochastic block model,'' 2015,
  http://arxiv.org/abs/1505.03772.

\bibitem{Mas14}
L.~Massouli{\'e}, ``Community detection thresholds and the weak {R}amanujan
  property,'' in \emph{Proc. ACM-SIAM Symp. Disc. Alg. (SODA)}, 2014, pp.
  694--703.

\bibitem{Mos12}
E.~Mossel, J.~Neeman, and A.~Sly, ``Stochastic block models and
  reconstruction,'' 2012, http://arxiv.org/abs/1202.1499.

\bibitem{Dec11}
A.~Decelle, F.~Krzakala, C.~Moore, and L.~Zdeborov{\'a}, ``Asymptotic analysis
  of the stochastic block model for modular networks and its algorithmic
  applications,'' \emph{Physical Review E}, vol.~84, no.~6, 2011.

\bibitem{Abb16}
E.~Abbe, A.~Bandeira, and G.~Hall, ``Exact recovery in the stochastic block
  model,'' \emph{IEEE Trans. Inf. Theory}, vol.~62, no.~1, pp. 471--487, Jan.
  2016.

\bibitem{Haj14}
B.~Hajek, Y.~Wu, and J.~Xu, ``Achieving exact cluster recovery threshold via
  semidefinite programming,'' 2014, http://arxiv.org/abs/1412.6156.

\bibitem{Abb15}
E.~Abbe and C.~Sandon, ``Community detection in general stochastic block
  models: Fundamental limits and efficient recovery algorithms,'' 2015,
  http://arxiv.org/abs/1503.00609.

\bibitem{Mos13}
E.~Mossel, J.~Neeman, and A.~Sly, ``Belief propagation, robust reconstruction,
  and optimal recovery of block models,'' 2013, http://arxiv.org/abs/1309.1380.

\bibitem{And15}
A.~Y. Zhang and H.~H. Zhou, ``Minimax rates of community detection in
  stochastic block models,'' 2015, http://arxiv.org/abs/1507.05313.

\bibitem{Mos15}
E.~Mossel and J.~Xu, ``Density evolution in the degree-correlated stochastic
  block model,'' 2015, http://arxiv.org/pdf/1509.03281v1.pdf.

\bibitem{Des15}
Y.~Deshpande, E.~Abbe, and A.~Montanari, ``Asymptotic mutual information for
  the two-groups stochastic block model,'' 2015,
  http://arxiv.org/abs/1507.08685.

\bibitem{Gue14}
O.~Gu\'edon and R.~Vershynin, ``Community detection in sparse networks via
  {G}rothendieck's inequality,'' 2014, community detection in sparse networks
  via Grothendieck's inequality.

\bibitem{Bou13}
S.~Boucheron, G.~Lugosi, and P.~Massart, \emph{Concentration Inequalities: A
  Nonasymptotic Theory of Independence}.\hskip 1em plus 0.5em minus 0.4em\relax
  OUP Oxford, 2013.

\end{thebibliography}

\end{document}